\documentclass{mn2e}
\input{epsf}

\title[The mass and anisotropy of four dSph galaxies]
{The mass and velocity anisotropy of the Carina, Fornax, Sculptor and Sextans dwarf spheroidal galaxies}

\author[E. L. {\L}okas]
    {Ewa L. {\L}okas
    \\
    Nicolaus Copernicus Astronomical Center, Bartycka 18, 00-716 Warsaw, Poland}
\begin{document}

\maketitle

\begin{abstract}
We model the large kinematic data sets for the four Milky Way dwarf spheroidal (dSph)
satellites: Carina, Fornax, Sculptor and Sextans, recently published by Walker et al.
The member stars are selected using a reliable
dynamical interloper removal scheme tested on simulated data. Our member selection is more
restrictive than the one based on metallicity indicators as it removes not only contamination due to
Milky Way stars but also the unbound stars from the tidal tails. We model the cleaned data sets by
adjusting the solutions of the Jeans equations to the profiles of the projected velocity dispersion and
kurtosis. The data are well reproduced by models where mass follows light and the best-fitting stellar orbits are
isotropic to weakly tangential, as expected from the tidal stirring scenario.
The Fornax dwarf, with more than 2400 member stars,
is a dSph galaxy with the most accurately determined mass to date: its $1\sigma$ error following from the sampling
errors of the velocity moments is below 5 percent.
With mass-to-light ratio of 97 solar units, Sextans seems to be the most dark matter dominated
of the four dSph galaxies.
\end{abstract}


\section{Introduction}

Dwarf spheroidal (dSph) galaxies of the Local Group are believed to be among the most dark matter dominated
objects in the universe. Their masses have important implications for the problem of the missing satellites.
The Carina, Fornax, Sculptor and Sextans dSph galaxies, all known for a few decades, are classic, well studied
examples of this family of stellar systems. Although generally believed to be strongly
dark matter dominated, their actual mass-to-light ratios are still very uncertain. While the
tidal origin of their large velocity dispersions is almost certainly ruled out, their formation scenarios are
still debatable.

The accuracy of mass determinations of dSph galaxies can only be improved by increasing the size of kinematic
samples used in the modelling. Recently, new measurements of radial velocities for the four dwarfs were published
by Walker, Mateo \& Olszewski (2008a), significantly increasing the number of stars with measured velocities
(by an order of magnitude in the case of Fornax). In this Letter we use these data to constrain the total mass
and velocity anisotropy of the dwarfs. Our method relies on the modelling of the velocity moments, the
dispersion and kurtosis of the line-of-sight velocity distribution by solutions of the Jeans equations
({\L}okas 2002; {\L}okas, Mamon \& Prada 2005). An underlying assumption of such an analysis is that in spite of
being tidally affected by the gravitational field of the Milky Way the dwarfs can still be approximated as
self-gravitating, virialized systems. An important first step in the analysis is to use a reliable procedure
for the removal of interlopers: the unbound stars from the Milky Way and the tidal tails of the dwarfs. We will
show that for all the dwarfs the data are well reproduced by models where mass traces light and the velocity
anisotropy parameter is constant with radius.

\begin{table*}
\caption{Adopted parameters of the dwarfs. }
\label{adopted}
\begin{center}
\begin{tabular}{lllll}
& Carina & Fornax & Sculptor & Sextans \\
\hline
centre (J2000)               & RA=$6^{\rm h} 41^{\rm m} 37^{\rm s}$  & RA=$2^{\rm h} 40^{\rm m} 04^{\rm s}$    & RA=$1^{\rm h} 00^{\rm m} 09^{\rm s}$   & RA=$10^{\rm h} 13^{\rm m} 03^{\rm s}$ \\
			     & Dec=$-50^{\circ}58'00''$              & Dec=$-34^{\circ}31'00''$                & Dec=$-33^{\circ}42'30''$               & Dec=$-1^{\circ}36'54''$               \\
distance modulus $(m-M)_0$   & $ 20.03 \pm 0.09$                     & $ 20.7 \pm 0.12$                        & $ 19.54 \pm 0.08$                      & $ 19.67 \pm 0.08$                     \\
distance $D$                 & $ 101 \pm 5$ kpc                      & $138 \pm 8$ kpc                         & $ 79 \pm 4 $ kpc                       & $ 86 \pm 4 $ kpc                      \\
apparent magnitude $m_V$     & $11.0 \pm 0.5$                        & $7.4 \pm 0.3$                           & $8.6 \pm 0.5$                          & $ 10.4 \pm 0.5$                       \\
absolute magnitude $M_V$     & $-9.03 \pm 0.59$                      & $ -13.3 \pm 0.42$                       & $ -10.94 \pm 0.58$                     & $ -9.27 \pm 0.58$                     \\
luminosity $L_V$             & $(3.50 \pm 1.35)\times 10^5 L_{\sun}$ & $(1.79 \pm 0.69) \times 10^7 L_{\sun}$  & $(2.03 \pm 0.79) \times 10^6 L_{\sun}$ & $(4.37 \pm 1.69) \times 10^5 L_{\sun}$ \\
S\'ersic radius $R_{\rm S}$  & 5.5 arcmin                            & 9.9 arcmin                              & 6.8 arcmin                             & 15.5 arcmin                           \\
S\'ersic parameter  $m$      & 1.0                                   & 1.0                                     & 1.0                                    & 1.0                                   \\
major axis PA                & $65^\circ$                            & $41^\circ$                              & $99^\circ$                             & $56^\circ$                            \\
\hline
\end{tabular}
\end{center}
\end{table*}

\begin{figure}
\begin{center}
    \leavevmode
    \epsfxsize=8.3cm
    \epsfbox[100 30 480 390]{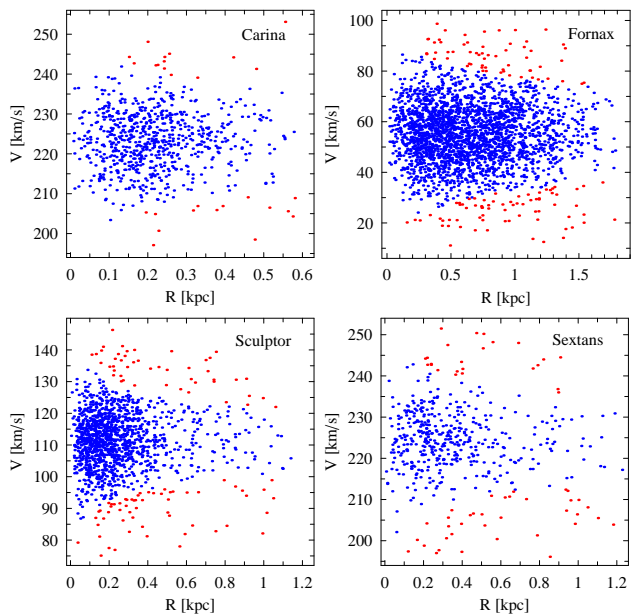}
\end{center}
\caption{The selection of member stars. Each point shows the heliocentric velocity of a star versus its
projected distance from
the centre of the dwarf. Blue dots correspond to the stars accepted by our iterative interloper removal
scheme in the final iteration. Red dots indicate the stars rejected by the procedure.}
\label{vr}
\end{figure}

All the assumptions adopted here have been verified by $N$-body simulations performed within the framework of the tidal
stirring scenario (Mayer et al. 2001). Such simulations show that dSph galaxies form effectively from initial stellar
disks embedded in dark matter haloes by tidal interactions. The products of such evolution have mass-to-light
profiles and anisotropy parameters almost constant with radius (Klimentowski et al. 2007; Mayer et al. 2007). In spite
of losing considerable fractions of mass by tidal forces, such objects are still gravitationally bound and their velocity
moments can be reliably reproduced by equilibrium models yielding realistic mass and anisotropy estimates
(Klimentowski et al. 2007; {\L}okas et al. 2008). Since
tidal tails in the immediate vicinity of the dwarfs are most likely oriented along the line of sight and are quite
dense (Klimentowski et al. 2008), we can however expect significant contamination of kinematic samples by unbound
stars. Still, such contamination can be effectively removed by an iterative procedure based on a maximum velocity available
to a given star (den Hartog \& Katgert 1996; Klimentowski et al. 2007; Wojtak et al. 2007).

\section{The data}

Figure~\ref{vr} presents the kinematic data sets for the four dwarfs from Walker et al. (2008a). In each panel
the dots correspond to heliocentric velocities of the stars as a function of their distance from the galaxy centre.
The positions of the centres and the distances of the dwarfs were adopted from Mateo (1998) except for Fornax,
where a newer estimate of the centre by Walcher et al.
(2003), which agrees well with the one of Coleman et al. (2005), was used. All these values are listed in
Table~\ref{adopted}. The velocity range shown in Fig.~\ref{vr} corresponds to a cut-off of about $4\sigma$ with
respect to the mean velocity of the galaxy, where $\sigma$ is the velocity dispersion in the centre, where
the contamination is low. Colour-coded in the Figure are the results of the interloper removal scheme
applied to the data.

The procedure, first proposed by den Hartog \& Katgert (1996) and applied to galaxy clusters,
relies on rejecting stars with velocity exceeding the maximum velocity available
to a star at a given projected distance from the centre. The maximum
velocity is estimated by assuming that the star is on a
circular orbit with velocity $v_{\rm cir} = [GM(r)/r]^{1/2}$ or falling freely in the galaxy's
potential with velocity $v_{\rm inf} = \sqrt{2}v_{\rm cir}$. The mass profile needed for the calculation of the
velocities is estimated in each iteration using the standard mass estimator $M_{VT}$ derived from the virial
theorem (Heisler, Tremaine \& Bahcall 1985). The procedure has been tested on mock data sets generated from
$N$-body simulations and shown to effectively remove most of unbound
stars while rejecting very few bound ones (Klimentowski et al. 2007; Wojtak et al. 2007). Note that the
efficiency of the procedure is not restricted to systems where mass follows light and it does not impose such
models (Klimentowski et al. 2007; S\'anchez-Conde et al. 2007).

The results of the
application of the procedure to the present data are illustrated in Fig.~\ref{vr} so that the blue dots show the
stars accepted in the final iteration while the red ones indicate those rejected by the procedure. Within the
velocity and distance ranges shown in the Figure we identified 700, 2445, 1281 and 412 members respectively
for Carina, Fornax, Sculptor and Sextans.  In the following
analysis we will only use these cleaned (blue) samples. We have verified that the procedure does not depend on the
initial velocity cut-off, i.e. identical final samples are obtained if the initial velocity cut-off is
$5\sigma$ or $3\sigma$ with respect to the galaxy mean.

Note that
our dynamical interloper removal scheme is more restrictive (i.e. it rejects more stars) than the method of sample
selection based on metallicity indicators discussed by Walker et al. (2008b). The latter is intended to identify
all the present as well as former stellar members of the dwarf so in principle it rejects only the contamination
from the Milky Way stars. On the other hand, our scheme rejects also the stars that were stripped from the dwarf due
to tidal interactions, which now populate the tidal tails and are no longer good tracers of the dwarf galaxy potential.

\begin{figure}
\begin{center}
    \leavevmode
    \epsfxsize=8.3cm
    \epsfbox[100 30 480 390]{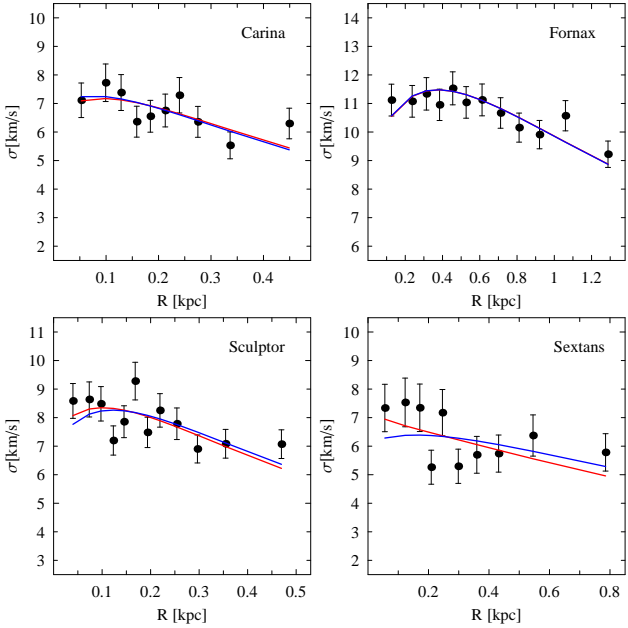}
\end{center}
\caption{Velocity dispersion profiles for the member stars in the four dSph galaxies. Errors indicate the
sampling errors of the dispersion. Red lines show the best-fitting profiles from fitting velocity dispersion
alone, blue ones from fitting velocity dispersion and kurtosis.}
\label{vd}
\end{figure}

\begin{figure}
\begin{center}
    \leavevmode
    \epsfxsize=8.3cm
    \epsfbox[100 30 480 390]{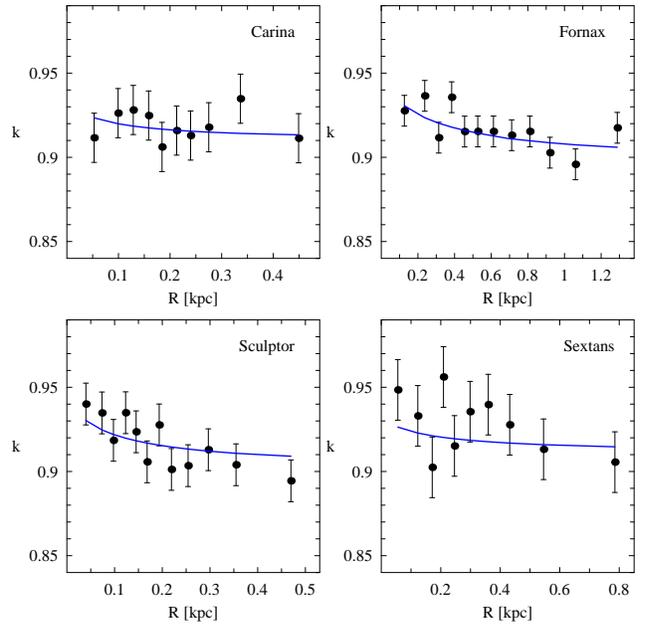}
\end{center}
\caption{The profiles of the kurtosis-like variable $k$ for the member stars in the four dSph galaxies.
Errors indicate the sampling errors. Blue lines show the best-fitting profiles from the joint fitting of the
velocity dispersion and kurtosis.}
\label{vk}
\end{figure}

In order to calculate the velocity moments we binned the data so that we had 70, 200, 100 and 40 stars per bin
respectively for Carina, Fornax, Sculptor and Sextans. From the data grouped in this way, the profiles of the
line-of-sight velocity dispersion
$\sigma(R)$ and kurtosis $\kappa(R)$ were calculated using standard unbiased estimators of these moments
(see {\L}okas et al. 2005).
The dispersion data points $\sigma(R)$ were assigned sampling errors of size $\sigma/\sqrt{2(n-1)}$
where $n$ is the number of stars per bin. Instead of kurtosis $\kappa$ we use the kurtosis-like
variable $k = (\log \kappa)^{1/10}$ which has a Gaussian sampling distribution contrary to
the kurtosis $\kappa$. The values of $k$ were obtained with the correction of the standard estimator
of kurtosis $K$ by the bias due to the low number of stars per bin so that $\kappa=3 K/C$ where $C= 2.75, 2.84,
2.89$ and 2.93 respectively for $n=40, 70, 100$ and 200. The sampling errors of $k$ are $0.018, 0.0146,
0.0124$ and 0.00914 for the same values of $n$.
These numerical values were estimated by Monte Carlo sampling of data from near-Gaussian distribution similar to
the one described in the Appendix of {\L}okas \& Mamon (2003).
The profiles of the velocity moments obtained for the data cleaned of interlopers (blue dots in Fig.~\ref{vr})
are plotted together with their sampling errors in
Figs.~\ref{vd} (velocity dispersion) and \ref{vk} (kurtosis). We can see that they are all slowly decreasing
with radius so our method of member selection leads to significantly different dispersion profiles than in the case
where no such dynamical scheme is applied and the profiles turn out to be flat (Walker et al. 2007).

\section{Results}

The data shown in Figs.~\ref{vd} and \ref{vk} were modelled by adjusting the projected solutions of the Jeans equations
(see {\L}okas 2002; {\L}okas et al. 2005) assuming that the mass distribution follows the light and the anisotropy
parameter $\beta$ is constant with radius. The shape parameters of the distribution of light
(i.e. the S\'ersic radius $R_{\rm S}$ and the S\'ersic parameter $m$) and the total apparent magnitude were adopted
from Irwin \& Hatzidimitriou (1995) and are listed in Table~\ref{adopted}. The Table also gives the total luminosity
for each galaxy, adjusted to the distances adopted here. The listed errors in total luminosity include the errors
in apparent magnitude as well as distance.

The lines plotted in Figs.~\ref{vd} and \ref{vk} show the best-fitting solutions of the Jeans equations. The red lines
(for Fornax not visible as overplotted by the blue line) correspond to the best
models when only the dispersion was fitted. The blue lines show the results when both velocity dispersion and
kurtosis profiles are fitted for each dwarf. In each case two parameters, the total mass $M$ and the anisotropy
parameter $\beta$ where adjusted. The best-fitting values of these parameters are listed in Table~\ref{fitted}
together with $1\sigma$ error bars following from the sampling errors of velocity moments. The
Table also gives the mass-to-light ratio with an error including both the inferred uncertainty of the mass and the
error in total luminosity.

\begin{figure}
\begin{center}
    \leavevmode
    \epsfxsize=8.3cm
    \epsfbox[7 0 390 385]{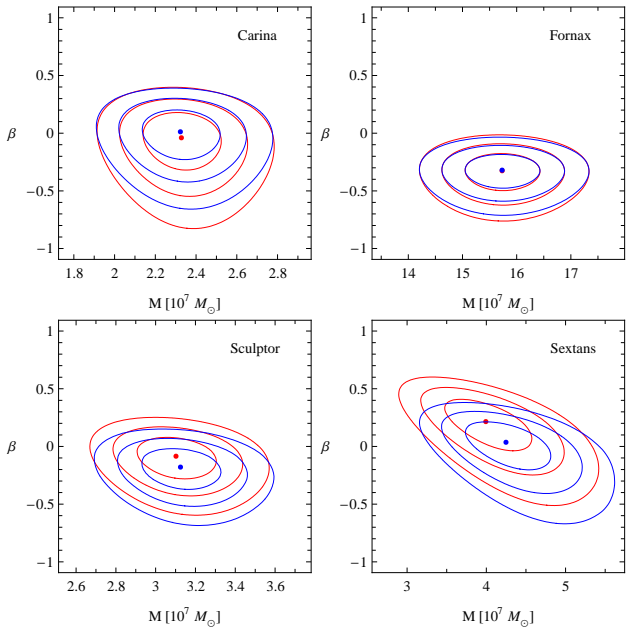}
\end{center}
\caption{The $1\sigma$, $2\sigma$ and $3\sigma$ probability contours showing constraints on the two fitted
parameters: the total mass $M$ and velocity anisotropy $\beta$ following from the sampling errors of
velocity moments. Red contours correspond to constraints from fitting the dispersion profile alone, while
blue ones to the case when both moments are fitted. Dots indicate the best-fitting parameters in each case.}
\label{cont}
\end{figure}

The obtained constraints on the parameters are further illustrated in Fig.~\ref{cont} which shows the $1\sigma$,
$2\sigma$ and $3\sigma$ probability contours in the $M$-$\beta$ parameter plane. Again, the red contours correspond
to the case when only the dispersion profiles were fitted, while the blue ones to the results of fitting both
velocity moments. The dots of corresponding colour mark the best-fitting parameters in both cases. For all dwarfs
including the kurtosis in the analysis tightens the constraints on anisotropy. Except for Fornax,
it also shifts a little the best-fitting parameters.

\begin{table*}
\begin{center}
\caption{Fitted parameters of the dwarfs. }
\begin{tabular}{lccccc}
galaxy & fitted & $\beta$     & $M[10^7 $M$_\odot]$   & $M/L_V[$M$_\odot/$L$_\odot]$ & $\chi^2/N$ \\
\hline
Carina   & $\sigma$        & $-0.04^{+0.24}_{-0.28}$ & $2.3 \pm 0.2$ & $67 \pm 31$ & $7.9/8$  \\

         & $\sigma+\kappa$ & $\ \; \; 0.01^{+0.19}_{-0.24}$ & $2.3 \pm 0.2$ & $66 \pm 31$ & $12.1/18$   \\ \\

Fornax   & $\sigma$        & $-0.33^{+0.15}_{-0.19}$ & $15.7 \pm 0.7$ & $8.8 \pm 3.8$ & $6.7/10$   \\

	 & $\sigma+\kappa$ & $-0.32^{+0.14}_{-0.17}$ & $15.7 \pm 0.7$ & $8.8 \pm 3.8$ & $17.7/22$ \\ \\

Sculptor & $\sigma$        & $-0.09^{+0.17}_{-0.20}$ & $3.1 \pm 0.2$ & $15.3 \pm 6.9$ & $14.6/10$  \\

	 & $\sigma+\kappa$ & $-0.18^{+0.16}_{-0.20}$ & $3.1 \pm 0.2$ & $15.4 \pm 7.0$ & $23.5/22$ \\ \\

Sextans  & $\sigma$        & $\ \; \; 0.22^{+0.18}_{-0.27}$ & $4.0 \pm 0.6$ & $91 \pm 49$ & $12.7/8$   \\

	 & $\sigma+\kappa$ & $\ \; \; 0.04^{+0.18}_{-0.25}$ & $4.2 \pm 0.6$ & $97 \pm 51$ & $23.9/18$ \\
\hline
\label{fitted}
\end{tabular}
\end{center}
\end{table*}

Our total mass estimates are much lower than the virial masses found by Walker et al. (2007) from the analysis
of velocity dispersion using similar
data sets, but with different member selection and assuming NFW dark matter profiles. As already mentioned, their less
restrictive member selection method leads to flat
velocity dispersion profiles which indeed can be reproduced by extended dark matter haloes. Our results agree
however with the masses they estimate for the inner parts of the dwarfs.

In order to illustrate the differences in our approaches to interloper rejection we performed a similar modelling
for Fornax as before but on a sample selected with the traditional constant velocity cut-off of $\pm 3 \sigma =
\pm 33$ km s$^{-1}$ with respect to the galaxy mean velocity. As expected, the velocity dispersion and kurtosis profiles
are now flat and constant $M/L$ models do not provide very good fits. When fitting only the dispersion the best-fitting
parameters are $M=(18.6 \pm 0.8) \times 10^7 $M$_\odot$ ($M/L_V=10.4 \pm 4.4 $M$_\odot/$L$_\odot$), $\beta=
-0.77^{+0.22}_{-0.27}$ with $\chi^2/N = 21.7/10$, when fitting both dispersion and kurtosis we find
$M=(18.4 \pm 0.8) \times 10^7 $M$_\odot$ ($M/L_V=10.3 \pm 4.4 $M$_\odot/$L$_\odot$), $\beta=
-0.50^{+0.15}_{-0.20}$ with $\chi^2/N = 58.1/22$. The differences with our estimates in Table~\ref{fitted} are not
very large, especially when both moments are fitted. This is due to the fact that contamination increases the
values of both dispersion and kurtosis while their dependence on anisotropy is different (see the discussion in {\L}okas
et al. 2008).

It may seem that the number of stars we rejected as interlopers (see Fig.~\ref{vr}) is large, especially for Fornax.
Note however, that significant contamination is expected for many of the dwarfs since for realistic, eccentric,
cosmologically motivated orbits their tidal tails are oriented approximately
towards the Milky Way (so towards the observer) for almost all positions along the orbit
except close to the pericentre, where the dwarfs spend only a small fraction of the orbital period
(Klimentowski et al. 2008).

\section{Discussion}

The new kinematic data sets for the Carina, Fornax, Sculptor and Sextans allowed us to determine the mass
of the dwarfs with unprecedented precision. In the case of Fornax, with the largest sample, the mass is
determined with better than 5 percent accuracy. This is due to the very small sampling errors of velocity
moments. The mass-to-light ratios are much worse constrained because of rather large errors in the determination
of luminosity. In spite of these, a dichotomy is seen among the studied dwarfs: while Fornax and Sculptor have
a mass-to-light of the order of 10 solar units, Carina and Sextans have significantly higher values of 66 and 97.
Although these high values still bear an error of about 50 percent, they cannot be reconciled with any stellar
population values thus signifying that these two dwarfs are highly dark matter dominated and similar in this
respect to the Draco dwarf ({\L}okas et al. 2005).

Our mass and mass-to-light determination for Carina agrees within errors with previous estimates obtained
from the total velocity dispersion by Mu\~noz et al. (2006) and Mateo et al. (1993) if we take into account their
different assumptions about the parameters of the luminosity distribution.
For Fornax the present estimates are somewhat lower than the previous ones (Walker et al. 2006a; Klimentowski et al.
2007; {\L}okas, Klimentowski \& Wojtak 2007). {\L}okas et al. (2007)
found $M/L$ of the order of 11 and isotropic orbits from a similar analysis as presented here but
based on $\sim200$ stars from Walker et al. (2006a), a number an order of magnitude smaller than presently available.
The anisotropy determined here is weakly tangential; note however that
although the orbits are still quite close to isotropic, isotropy is excluded at $3\sigma$ level.

In the case of Sculptor our $M/L$ estimate is an order of magnitude lower than the most recent one by
Battaglia et al. (2008) who find $M/L = 158$ at the distance of 1.8 kpc, but agrees well with earlier estimates by
Queloz, Dubath \& Pasquini (1995) and Armandroff \& Da Costa (1986). The reason for this large discrepancy lies
in the selection of the data as well as in models used. Note that the fraction of stars discarded by our interloper
removal scheme (Fig.~\ref{vr}) is large in the case of Sculptor so the data are highly contaminated leading to larger
velocity dispersion found by Battaglia et al. The contamination probably comes from the tidal tails of the dwarf:
as discussed by Coleman, Da Costa \& Bland-Hawthorn (2005) the tails are not visible in photometric studies so they
must be oriented along the line of sight thus contributing significantly to the kinematics (Klimentowski et al. 2007).
In addition, the data sample of Battaglia et al. extends to larger distances from the centre of the dwarf than the
one studied here, making the contamination even more probable. This manifests itself in the flat rather than
decreasing velocity dispersion profile suggesting the presence of an extended dark matter halo with a slowly decreasing
density profile that can indeed lead to the inference of large $M/L$ values at large distances.

The mass-to-light ratio for Sextans estimated from the total velocity dispersion by Hargreaves et al. (1994)
is 94 solar units (when translated to our adopted luminosity) in perfect agreement with our value of 97. These
values are significantly higher than those estimated by Suntzeff et al. (1993) who find $M/L$ of the order
of $30-54$. We thus confirm that this dwarf is indeed very strongly dark matter dominated.
Our determined mass also agrees with the rough estimate by Kleyna et al. (2004) who found
the mass in the range $3 \times 10^7 - 1.5 \times 10^8 $M$_\odot$ assuming isotropic orbits in the Jeans equation.
In agreement with Walker et al. (2006b) we find however no evidence for the presence of a cold kinematic core in Sextans
claimed by Kleyna et al. (2004).

We conclude that for all four dwarfs the data are satisfactorily reproduced by models where mass traces light and
the anisotropy is close to zero or weakly tangential. Both these findings yield support to the formation
scenarios based on the so-called tidal stirring model (Mayer et al. 2001). The model predicts that an extended
dark matter halo initially present in a dwarf galaxy is strongly stripped due to tidal interactions
with the Milky Way (Kazantzidis et al. 2004)
and the final product we see today should have dark matter distributed in a similar way as the stars
up to distances where the tidal tails start to be visible (Klimentowski et al. 2007; Mayer et al. 2007).
The stellar orbits are expected to be weakly tangential, e.g. in the simulation of Klimentowski et al. (2007)
the final stage has mean $\beta=-0.13$ (see also their fig. 4), of the same order as the values estimated here.
The tidal scenario also predicts that some remnant rotation could be expected in the dwarfs at present,
however, contrary to the previous claims e.g. for Carina (Mu\~noz et al. 2006) and Sculptor (Battaglia et al. 2008)
no significant rotation is found in the data studied here.

\section*{Acknowledgements}

I am grateful to M. Walker et al. for providing
the kinematic data for the dwarfs in convenient electronic form. This research was partially supported by the
Polish Ministry of Science and Higher Education under grant N N203 0253 33.

\end{document}